\documentclass[preprint,showpacs,titlepage,aps,prd,
tightenlines,
amsmath,byrevtex,nofootinbib]{revtex4}

\usepackage{graphicx}

\newcommand{\GeV}{{\text{GeV}}}
\newcommand{\best}{{\text{best}}}
\newcommand{\rel}{{\text{rel}}}
\newcommand{\muBG}{{\mu \text{BG}}}

\newcommand{\BG}{{\text{BG}}}
\newcommand{\rec}{{\text{rec}}}

\begin{document}

\preprint{KEK-TH-964}
\preprint{hep-ph/0406073}

\title{Determination of $\theta_{23}$ in Long-Baseline 
Neutrino Oscillation Experiments with Three-Flavor Mixing Effects}

\author{Hisakazu Minakata}
\email{E-mail: minakata@phys.metro-u.ac.jp}
\author{Masashi Sonoyama}
\email{E-mail: m-sono@phys.metro-u.ac.jp}
\affiliation{Department of Physics, Tokyo Metropolitan University,
Hachioji, Tokyo 192-0397, Japan}
\author{Hiroaki Sugiyama}
\email{E-mail: hiroaki@post.kek.jp}
\affiliation{Theory Group, KEK, Tsukuba, Ibaraki 305-0801, Japan}

\date{\today}

\vglue 1.4cm
\begin{abstract}

We examine accuracy of $\theta_{23}$ determination in future
long-baseline (LBL) $\nu_{\mu}$ disappearance experiments in the 
three-flavor mixing scheme of neutrinos. 
Despite that the error of $\sin^2{2\theta_{23}}$ is indeed a few\% level 
at around the maximal mixing, 
we show that the error of physics variable $s^2_{23}$ is large, 
$\delta(s^2_{23})/s^2_{23} \simeq 10 \mbox{-} 20\%$, 
depending upon regions of $\theta_{23}$. 
The errors are severely affected by the octant degeneracy of 
$\theta_{23}$, and $\delta(s^2_{23})$ is largely amplified by the 
Jacobian factor relating these two variables in a region near to the 
maximal mixing. 
The errors are also affected by the uncertainty due to unknown
value of $\theta_{13}$; $\delta(s^2_{23})$ is doubled at off maximal 
in the second octant of $\theta_{23}$ where the effect is largest.
To overcome this problem, we discuss combined analysis
with $\nu_e$ appearance measurement in LBL experiments, 
or with reactor measurement of $\theta_{13}$.
For possible relevance of sub-leading effects even in the 
next-generation LBL experiments, we give a self-contained 
derivation of the survival probability to the next to leading order in
$s^2_{13}$ and $\Delta m^2_{21} / \Delta m^2_{31}$.
\end{abstract}


\maketitle


\section{Introduction}
\label{introduction} 

At several years since the pioneering discovery in atmospheric neutrino 
observation by Super-Kamiokande (SK)~\cite{SKatm}, 
we now have compelling evidences for neutrino oscillation. 
The direct confirmation of oscillation of atmospheric neutrinos 
comes from the K2K experiment using man-made neutrino beam 
\cite{K2K} which confirmed its existence at 
3.9$\sigma$ CL \cite{K2K_new}. 
The neutrino oscillation with the solar $\Delta m^2$ scale is beautifully 
observed by the KamLAND experiment 
\cite{KamLAND,KamLAND_new}. 
By observing deficit in reactor antineutrino flux it pinned down the 
large-mixing-angle (LMA) solution of the solar neutrino problem based on  
the Mikheyev-Smirnov-Wolfenstein (MSW) mechanism~\cite{wolfenstein,MSW}.
It has thereby settled the long-standing solar neutrino problem \cite{BD}, 
whose origin was identified as due to the neutrino flavor 
transformation by SNO~\cite{SNO}.
Long awaited confirmation of the oscillatory behavior is now available 
by the three experiments, SK~\cite{SKatm_new}, K2K \cite{K2K_new}, and 
KamLAND~\cite{KamLAND_new}. 
Thus, we already know roughly the structure
of lepton flavor mixing in the (1-2) and (2-3) sectors of the
Maki-Nakagawa-Sakata (MNS) matrix~\cite{MNS}.

We recognize that there are two small (or possibly vanishing)
quantities in the lepton mixing matrix, $\theta_{13}$ and
deviation of $\theta_{23}$ from $\pi/4$.
The former is constrained mainly by the reactor experiments~\cite{CHOOZ} 
to be $\sin^2{2\theta_{13}} \leq 0.15$ at 90\%~CL 
by a global analysis~\cite{global}, while the latter is bounded as 
$0 \leq 1- \sin^2{2\theta_{23}} \leq 0.08$ at 90\%~CL
by the recent reanalysis of atmospheric neutrino data by
the SK group~\cite{kearns}. 
Possible correlation between these two small quantities may imply 
hints for the yet unknown discipline that was used by nature to
design the structure of lepton flavor mixing. Some symmetries 
have been discussed by which the maximal $\theta_{23}$ 
is correlated with vanishing $\theta_{13}$ in the symmetry limit 
\cite{symmetry}. Furthermore, the question of how large is the 
deviation of $\theta_{23}$ from $\pi/4$ is discussed from a 
variety of viewpoints \cite{deviation}.
Therefore, it is important to determine the value of
not only $\theta_{13}$ but also $\theta_{23}$ very precisely.

How accurately $\theta_{23}$ can be measured experimentally? 
It appears that measurement of disappearance probability of 
$\nu_{\mu}$ at 
around the first oscillation maximum has highest sensitivity to 
$\sin^2{2\theta_{23}}$.\footnote{
The statement applies also to the other large mixing angle, 
$\theta_{12}$, if $\bar{\nu}_e$ from reactors is used~\cite{MNTZ}. 
In this case, the accuracy for $\sin^2{2\theta_{12}}$ measurement 
translates into the one of $\sin^2{\theta_{12}}$ without suffering 
from the problems (apart from the third one) mentioned below. 
}
In the J-PARC $\rightarrow$ Super-Kamiokande 
(JPARC-SK) experiment, for example, $\sin^2{2\theta_{23}}$ 
is expected to be determined to 1\% accuracies \cite{JPARC}. 
Given such an enormous accuracy, it seems that there is 
not much to add because it looks like the best thinkable 
accuracy of mixing angle determination in neutrino experiments.
Unfortunately, it is not quite true. 
We discuss three relevant and mutually related issues in this paper 
which prevent an accurate determination of $\theta_{23}$, in particular, 
its small deviation from the maximal. 
The points we discuss are: 

\begin{itemize}

\item

a large Jacobian associated with transformation of the variable from 
$\sin^2{2\theta_{23}}$ to $s^2_{23}$ at around the maximal mixing

\item
the octant ambiguity of $\theta_{23}$ 
(two solutions of $\theta_{23}$ larger or smaller than $\pi/4$) 

\item

effects of three-flavor mixing, in particular of non-vanishing $\theta_{13}$

\end{itemize}

\noindent
While the first point is  ``kinematical'' in nature, the second 
and the third points reflect the three-flavor structure of lepton mixing. 
Although each of these points was discussed in various occasions in 
the past, we believe that they are never discussed in a coherent fashion 
in the context of precision measurement of $\theta_{23}$ by 
long-baseline (LBL) experiments. 
The estimation of errors of $\theta_{23}$ determination is also done 
by taking account of three-flavor mixing effects \cite{AHKSW,munich10}
with consistent results with ours. 
Yet, they offer neither understanding of the reasons behind the 
estimated large errors in the similar depth as we provide, 
nor ways of improvement by combining with e.g., reactor measurement.

Importance of experimental determination of $\theta_{13}$ is
now well recognized in the community by itself and 
as a door to exploration of leptonic CP violation.
Strategies for determining the remaining oscillation parameters
are developed with use of accelerator superbeam~\cite{superbeam} and
reactor neutrinos~\cite{MSYIS}, which followed by a series of
feasible experimental programs%
~\cite{JPARC,NuMI,SPL,reactor_white}.
The accuracies of $\theta_{13}$ determination in the LBL and
the reactor experiments are expected to reach to 
$\sin^2{\theta_{13}} \simeq 2.5 \times 10^{-3}$.
We feel that the strategy for similar accurate determination of
$\theta_{23}$ is relatively less well developed, and it is
the purpose of this paper to trigger such efforts.

By indicating that the three-flavor nature of
the oscillation probability already manifests itself
in the next generation LBL experiments, 
our discussion in this paper might have an implication to future 
neutrino oscillation research, suggesting a possible direction. 
The next to leading order corrections may be important due to 
the larger ratio of 
$\epsilon \equiv \Delta m^2_{21} / \Delta m^2_{31} \simeq 1/30$
due to a larger $\Delta m^2_{21}$ \cite{KamLAND_new}, 
and possible large value of $\theta_{13}$ close to the CHOOZ bound.
Under such hope, we present in Appendix A a derivation of the
full expression of survival probability
$P(\nu_{\mu} \rightarrow \nu_{\mu})$
to the next to leading order in $\epsilon$ and $s^2_{13}$, 
which is used in our analyses throughout this paper.
See also \cite{perturbative} for derivation of such formulas under the similar spirit.

\section{Why is accurate determination of $\theta_{23}$ so difficult?}
\label{whatsmore}

In this section we present pedagogical description of the 
three reasons for difficulty of accurate determination of $\theta_{23}$ 
that were mentioned in Sec.~\ref{introduction}. 
The readers who are familiar to them are advised to skip this section 
and go directly to Sec.~\ref{analysis}.

\subsection{Problems of large Jacobian at around the maximal mixing}
\label{Jacobian}

The ultimate purpose of the whole experimental activities 
which aim at measuring neutrino mixing parameters is to 
understand physics of lepton flavor mixing.
For this purpose precise determination of the angles or the 
sine of angles are required. 

The $\nu_{\mu}$ disappearance measurement is sensitive to 
$\sin^2{2\theta_{23}}$, not $s^2_{23}$, which is unfortunate, 
but is actually the case.
It inherently makes accuracy of determination of 
$\theta_{23}$ much worse than it is achieved for $\sin^2{2\theta_{23}}$ 
at around the maximal mixing. Assuming that the true value of 
$\theta_{23}$ is $\pi/4$, for example, an accurate measurement of 
$\sin^2{2\theta_{23}}$ with error of 1\% level translates into large 
error of $\pm 10$\% level for $s^2_{23}$.
It is simply because the Jacobian for the transformation of the variables 
\begin{eqnarray}
\frac{\delta(s_{23}^2)}{ \delta (\sin^2{2\theta_{23}}) } \simeq
\frac{d s^2_{23}}{d \sin^2{2\theta_{23}}} = 
\frac{1}{4 \cos{2\theta_{23}}},  
\label{jacobian}
\end{eqnarray} 
are large at around $\theta_{23}=\pi/4$, 
by which a tiny error obtained for $\sin^2{2\theta_{23}}$, 
denoted as $\delta (\sin^2{2\theta_{23}})$ in (\ref{jacobian}),  
can be translated into a large error $\delta (s_{23}^2)$ for $s_{23}^2$.
It is what happens in Fig.~\ref{ser_30_theta} in Sec.~\ref{analysis}.

We emphasize that use of $s^2_{23}$, not 
$\sin^2{2\theta_{23}}$, is inevitable in the error analysis based on 
three flavor mixing because the $\nu_{\mu}$ survival probability itself 
cannot be expressed in terms of $\sin^2{2\theta_{23}}$ only 
once next to the leading order correction is taken into account.
See (\ref{Pvac}) below and (\ref{survivalP}) for the probabilities 
in vacuum and in matter, respectively.
It is even more so if we combine $\nu_e$ appearance measurement 
into our analysis, as will be done in Sec.~\ref{app+react}, 
because the dominant term in the $\nu_e$ appearance probability 
depends upon $\theta_{23}$ through $s^2_{23}$.

\subsection{Octant ambiguity of $\theta_{23}$}
\label{degeneracy}

\begin{figure}[tbp]
\begin{center}
\includegraphics[width=0.7\textwidth]{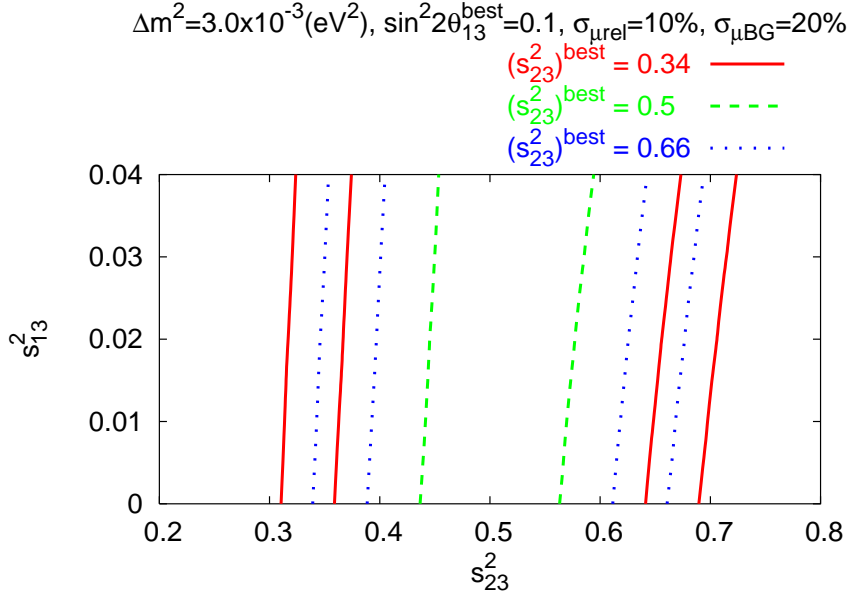}
\end{center}
\caption{
90\%~CL allowed regions are plotted in $s^2_{23} \mbox{-} s^2_{13}$ 
space by assuming that the best fit values given by nature are 
[$(s_{23}^2)^\best$, $(s_{13}^2)^\best$] = 
[0.34, 0.026] (red solid curve), [0.5, 0.026] (green dashed curve), and 
[0.66, 0.026] (blue dotted curve);
Note that $s_{23}^2 = 0.34$ and 0.66 correspond to
$\sin^2{2\theta_{23}} = 0.9$ which is the lowest edge of the current allowed region 
and $s_{13}^2 = 0.026$ corresponds to $\sin^2{2\theta_{13}} = 0.1$. 
The systematic errors are assumed to be  
($\sigma_{\mu \rel}$, $\sigma_{\muBG}$) = (10\%, 20\%).
%
}
\label{th-a_reg}
\end{figure}

The dominant term in $\nu_{\mu}$ survival probability, and hence 
the disappearance measurement, is sensitive primarily to $\sin^2{2\theta_{23}}$.
Therefore, there exist two solutions of $\theta_{23}$ 
for a given value of the survival probability 
$P(\nu_{\mu} \rightarrow \nu_{\mu})$, 
the octant ambiguity of $\theta_{23}$.\footnote{
It was recognized since sometime ago that there exists multiple solutions 
of mixing parameters, $\theta_{13}$, $\theta_{23}$ and $\delta$,
for a given set of measured oscillation probabilities. It is called 
the problem of parameter degeneracy 
\cite{Burguet-C,MNjhep01,octant,BMW1,KMN02,MNP2}, and the 
octant ambiguity of $\theta_{23}$ is only a part of it. 
The nature of the degeneracy may be characterized as 
the intrinsic degeneracy of $\theta_{13}$ and $\delta$ \cite{Burguet-C}, 
which is duplicated by 
the unknown sign of $\Delta m^2_{31}$ \cite{MNjhep01} and 
the octant ambiguity of $\theta_{23}$ for a given $\sin{2\theta_{23}}$ 
\cite{octant}. 
The octant ambiguity may be the most difficult one to resolve among 
the eight-fold parameter degeneracies unless it is combined with 
either reactor \cite{MSYIS} or silver channel \cite{silver} measurement.
}
%
In Fig.~\ref{th-a_reg} we present a contour of 90\%~CL allowed 
region in $s^2_{23} \mbox{-} s^2_{13}$  space by 5 years measurement at 
JPARC-SK\@ based on the analysis in Sec.~\ref{analysis} whose details 
will be explained in Appendix B. 

It should be noticed in Fig.~\ref{th-a_reg} that there exist  
``clone'' solutions, which arise as a consequence of the degeneracy 
of $\theta_{23}$ solutions~\cite{octant}. 
Namely, with an assumed true value $s^2_{23}=0.34$, for example, 
the measurement allows two regions as solutions, one around the original 
value of $s_{23}$ in the first octant and the other one around
$s^2_{23}=0.66$ which is the 
reflected point with respect to $\theta_{23}=\pi/4$ in the second octant. 
Of course, it does not produce additional experimental error of $s^2_{23}$ 
as far as two regions are separated, though ambiguity exists of which 
region contains the true value of $\theta_{23}$. 
But once the two allowed regions start to merge 
(which takes place near the maximal $\theta_{23}$), 
it does produce additional error of $s^2_{23}$ as exhibited in the spike 
in Fig.~\ref{ser_30_theta}.
The similar figure as Fig.~\ref{th-a_reg} indicating 
$\theta_{23} \mbox{-} \theta_{13}$ correlation and clone solution is used in a 
slightly different context in \cite{gab-gouvea}.

\subsection{Three-flavor mixing effects in the disappearance 
measurement of $\theta_{23}$}
\label{3flavor}

While the effect of three-flavor mixing gives relatively mild influence 
to the accuracy of $\theta_{23}$ measurement compared to the 
above two effects, it is nonetheless non-negligible. 
Let us explain the key point by taking a very simple case 
using the approximation of vacuum oscillation and 
small deviation from maximality of $\theta_{23}$.
Though simplified, the model captures some relevant features 
possessed by the more realistic cases. 

We use, throughout this paper, the standard notation of 
the MNS matrix~\cite{PDG},
\begin{eqnarray}
U=\left[
\begin{array}{ccc}
c_{12}c_{13} & s_{12}c_{13} &   s_{13}e^{-i\delta}\\
-s_{12}c_{23}-c_{12}s_{23}s_{13}e^{i\delta} &
c_{12}c_{23}-s_{12}s_{23}s_{13}e^{i\delta} & s_{23}c_{13}\\
s_{12}s_{23}-c_{12}c_{23}s_{13}e^{i\delta} &
-c_{12}s_{23}-s_{12}c_{23}s_{13}e^{i\delta} & c_{23}c_{13}\\
\end{array}
\right],
\label{MNSmatrix}
\end{eqnarray}
where  $c_{ij}$ and $s_{ij}$ ($i, j = 1 \mbox{-} 3$) imply 
$\cos{\theta_{ij}}$ and $\sin{\theta_{ij}}$,  
respectively. The mass squared difference of neutrinos 
is defined as $\Delta m^2_{ij} \equiv m^2_i - m^2_j$ where
$m_i$ is the eigenvalue of the $i$th mass-eigenstate.
Then, the disappearance probability 
$1-P(\nu_{\mu} \rightarrow \nu_{\mu})$ in vacuum is given by 
\begin{eqnarray}
1- P(\nu_{\mu} \rightarrow \nu_{\mu}) & = &
\sin^2 2\theta_{23}\sin^2\Bigl(\frac{\Delta m^2_{31} L}{4E} \Bigr)
+ 4s_{13}^2s_{23}^2(s_{23}^2-c_{23}^2)\sin^2
\Bigl(\frac{\Delta m^2_{31} L}{4E} \Bigr)
\nonumber \\
& - & c_{12}^2\sin^22\theta_{23}
\Bigl(\frac{\Delta m_{21}^2 L}{4E}\Bigr)
\sin\Bigl(\frac{\Delta m_{31}^2 L}{2E}\Bigr) .
\label{Pvac}
\end{eqnarray}
In the right-hand side of (\ref{Pvac}),
the first term represents major contribution 
which survives in the two-flavor or the single mass-scale 
dominance approximation \cite{one-mass}. 
The second and the third terms are small corrections due to 
non-vanishing $s^2_{13}$ and 
$\epsilon \equiv \Delta m^2_{21} / \Delta m^2_{31}$, respectively,
which give effects of the three-flavor mixing.
It is obvious that the second term vanishes 
for the maximal $\theta_{23}$, while the third term vanishes 
at the oscillation maximum, $\Delta m_{31}^2 L/2E= \pi$.

Suppose that we perform a hypothetical experiment at a monochromatic 
energy close (but not identical) to the oscillation maximum,
$\sin{(\Delta m_{31}^2 L/2E)} \ll 1$. 
We assume that deviation from maximality of $\theta_{23}$ is small, 
and parametrize it by $\eta$ defined as 
$\eta \equiv \cos{2\theta_{23}}$ 
(so that $1-\sin^2{2\theta_{23}} = \eta^2 \ll 1$). 
Using the approximation $\sin{(\Delta m_{31}^2 L/4E)} = 1$ 
we obtain 
\begin{eqnarray}
P(\nu_{\mu} \rightarrow \nu_{\mu}) = 
\eta (\eta + 2 s^2_{13}) + 
\frac{\pi}{2} c^2_{12} \epsilon
\sin\left( \frac{\Delta m^2_{31} L}{2E} \right) .
\label{PvacOM}
\end{eqnarray}
Notice that the linear term in $\eta$ in (\ref{PvacOM}) 
distinguishes between $\theta_{23}$ in the first and the 
second octants because $\eta$ changes sign depending upon 
which octant it lives. 
%
%
The equation (\ref{PvacOM}) clearly indicates
that the larger value of $s_{13}^2$ gives
the smaller (larger) value of $\eta$
in first (second) octant of $\theta_{23}$
for a given $P(\nu_\mu\to\nu_\mu)$.
It is interesting to observe that the feature of 
$\theta_{23}$-$\theta_{13}$ correlation is in fact shared in 
Fig.~\ref{th-a_reg}.

Therefore, what the disappearance measurement actually determines
is not $\sin^2{2\theta_{23}}$ but allowed regions in 
$s^2_{23} \mbox{-} s^2_{13}$ space.
The feature leads to larger errors of $\theta_{23}$ in 
three-flavor analysis compared to those of two-flavor analysis. 
It is notable that the effect of $s_{13}^2$ is slightly 
more significant in the second octant of $\theta_{23}$.
Note also the tendency of departure from the maximal $\theta_{23}$ 
due to $\theta_{13}$ is smaller (larger) 
at larger $\theta_{13}$ in the first (second) octant of $\theta_{23}$. 
These properties are the key to fuller understanding of the features 
of the errors of $\theta_{23}$ to be discussed in the next section. 
The problem of $\theta_{13}$ dependence of accuracy of $\theta_{23}$
determination has been addressed, e.g., in \cite{barger_etal}.

Another notable feature in (\ref{PvacOM}) is that 
the solar mass scale correction can come into play a role.
It can change the depth of the dip of $P(\nu_{\mu} \rightarrow \nu_{\mu})$ 
and therefore may affect the value of $\eta$.

\section{Disappearance measurement of $\theta_{23}$ 
and three-flavor mixing effect}
\label{analysis}

In this section, we fully analyze the sensitivity of $\theta_{23}$ 
by $\nu_{\mu}$ disappearance measurement by taking account 
of the three-flavor mixing effects. 
To make our discussion as concrete as possible, we take the 
particular experiment, the JPARC-SK experiment~\cite{JPARC}. 
While the Jacobian and the degeneracy effects are universal to 
any experiments, the three-favor effect is most prominent if 
the systematic uncertainties of the experiment are small 
enough to allow measurement of $\sin^2{2\theta_{23}}$ to 
a few~\% level accuracies. 
We describe in detail the statistical method used in our analysis in 
Appendices B and C so that we can concentrate on physics 
discussion in the main text. 
Through the course of formulating our analysis procedure, 
we will try to reproduce the results of sensitivity analyses 
by the experimental group. (See Fig.~\ref{LOI_comp}.)

We take 22.5 kt as the fiducial volume of SK and assume, 
throughout this paper,  
5 years running of neutrino mode $\nu_{\mu}$ disappearance 
measurement, as expected in its phase I defined in LOI\@.  
We examine two cases of systematic errors, 
($\sigma_{\mu \rel}$, $\sigma_{\muBG}$) = (10\%, 20\%) and 
(5\%, 10\%). The former numbers, which are based on their experience 
in K2K, can be too pessimistic but are quoted in LOI, while the 
latter may be a reasonable goal to be reached in phase~I of the 
J-PARC neutrino project. 
To make the goal and to achieve even smaller errors the group plans to 
build an intermediate detector at $\sim$~2km from J-PARC\@. 
We use, unless otherwise stated, 
$\Delta m_{31}^2 = 3\times 10^{-3}$eV$^2$, 
$\Delta m_{21}^2 = 7.3\times 10^{-5}$eV$^2$ 
and $s_{12}^2 = 0.315$. We take the earth matter density 
$\rho = 2.3\,\mbox{g}\cdot\mbox{cm}^{-3}$ based on the estimation 
quoted in~\cite{koike-sato}.

We assume that $\Delta m^2_{31} > 0$ in our analysis.
We have checked that the error $\delta(\sin^2{2\theta_{23}})$ 
slightly increases in the case of negative $\Delta m^2_{31}$ 
but only by about 0.001 in the allowed region of $\theta_{23}^\best$. 
In this paper, the quantity with superscript ``best'' always implies the 
best fit value given by nature. 
We have also checked that our results do not change in any appreciable 
manner by taking $\Delta m^2_{21}$ at the highest end of the 
LMA-I parameter, $\Delta m^2_{21} = 10^{-4} \mbox{eV}^2$. 
Therefore, we do not discuss further about the possible 
effects caused by flipping the sign of $\Delta m^2_{31}$ and 
by raising the solar $\Delta m^2_{21}$.

\subsection{Accuracy of determination of $s^2_{23}$ vs.\ that of $\sin^2{2\theta_{23}}$}

\begin{figure}[tbp]
\begin{center}
\includegraphics[width=0.6\textwidth]{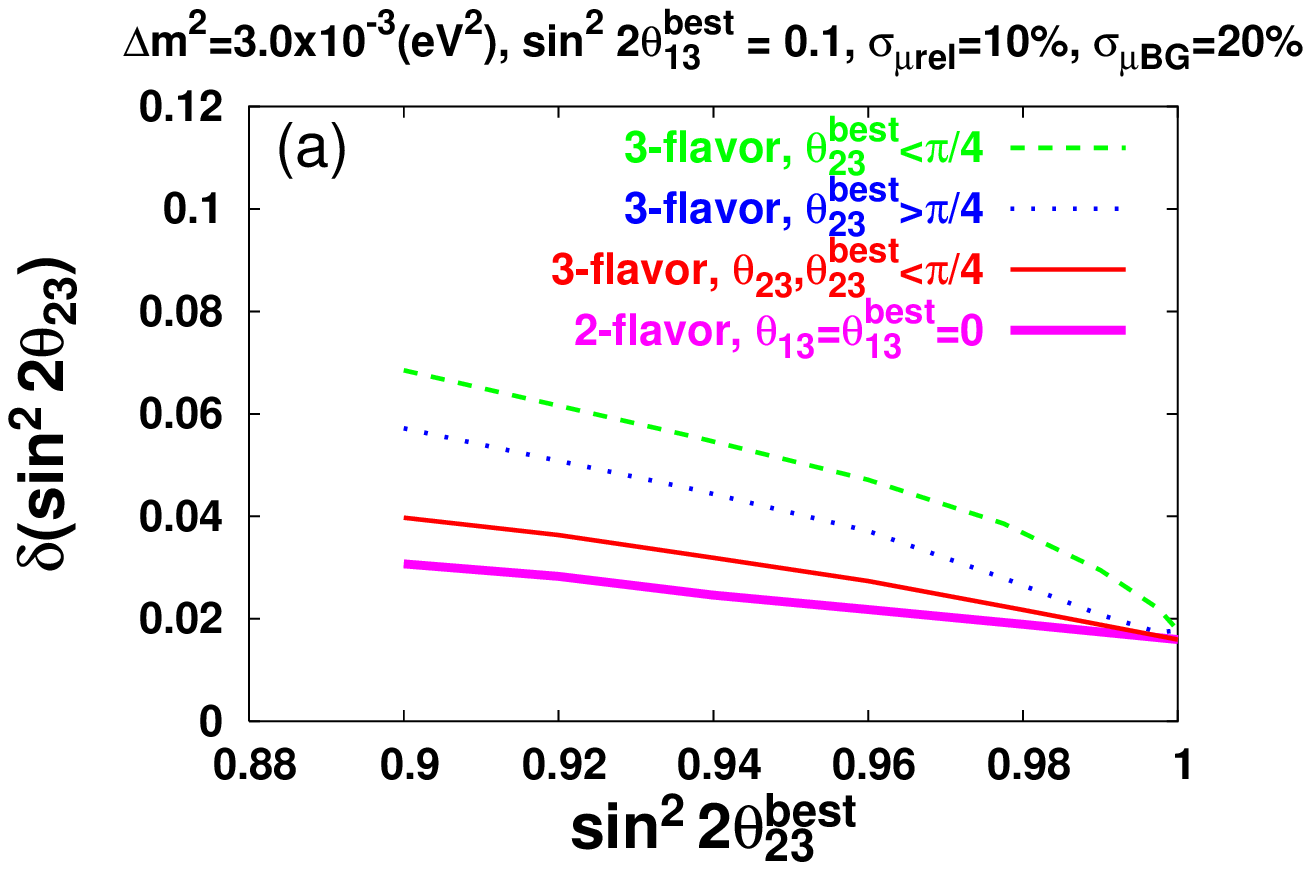}
\includegraphics[width=0.6\textwidth]{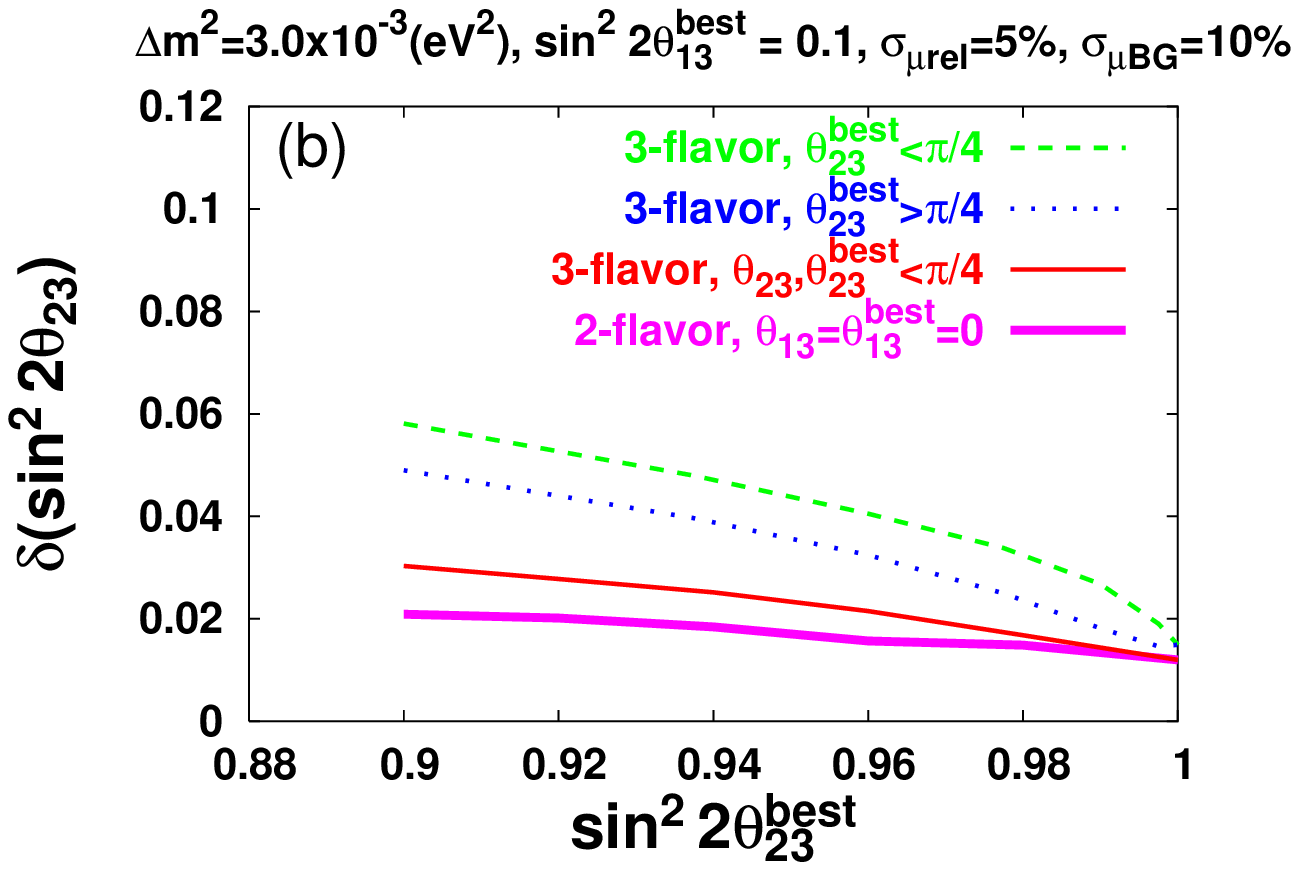}
\end{center}
\caption{The expected accuracies of determination of 
$\sin^2{2\theta_{23}}$ at 90\% CL obtained by the three-flavor analysis 
are compared with that of our two-flavor analysis
($\theta_{13}=\theta_{13}^\best =0$)
which is indicated by pink thick-solid line.
The red thin-solid line is the result obtained in the analysis assuming 
that the $\theta_{23}$ is in the first octant, $0 \leq \theta_{23} \leq \pi/4$. 
The green dashed and blue dotted lines are for input values of 
$\theta_{23}^\best$ in the first and the second octants, respectively, 
while allowing the output $\theta_{23}$ in the full range, 
$0 \leq \theta_{23} \leq \pi/2$.
Fig.~\ref{ser_30_2theta}a and \ref{ser_30_2theta}b are 
with systematic errors 
($\sigma_{\mu \rel}$, $\sigma_{\muBG}$) = (10\%, 20\%) and 
(5\%, 10\%), respectively.
}
\label{ser_30_2theta}
\end{figure}

We compare the accuracies of determination of $\sin^2{2\theta_{23}}$ 
and $s^2_{23}$ to illuminate how the problem of Jacobian and the 
degeneracy affect the accuracy of  $\theta_{23}$ determination. 
We show in Fig.~\ref{ser_30_2theta} the errors expected for 
determination of $\sin^2{2\theta_{23}}$ as a function 
of $\sin^2{2\theta_{23}^\best}$.
In Fig.~\ref{ser_30_2theta}a and Fig.~\ref{ser_30_2theta}b, 
the systematic errors are taken as 
($\sigma_{\mu \rel}$, $\sigma_{\muBG}$) = (10\%, 20\%) and 
(5\%, 10\%), respectively. 
The figure includes not only the result of the three-flavor analysis 
with the particular nonzero input value of 
$\theta_{13}$, $\sin^2{2\theta_{13}^\best} = 0.1$,  
but also that of the two-flavor analysis assuming 
$\theta_{13}=\theta_{13}^\best =0$.
We note that the results depend on input values of $\theta_{13}^\best$ 
but only mildly; If we take $\sin^2{2\theta_{13}^\best} = 0$ in the three flavor 
analysis, the uncertainty of 
$\sin^2{2\theta_{23}}$ decreases only by $\sim$5\%.

We call special attention to the distinction between the lower solid lines,  
and the upper dashed and dotted lines in the three-flavor results in 
Fig.~\ref{ser_30_2theta}. 
Namely, the red thin-solid and the pink thick-solid line are 
obtained under the assumption that $\theta_{23}$ is in the first octant, 
$0 \leq \theta_{23} \leq \pi/4$, by which the octant degeneracy is 
switched off by hand. This procedure is adopted, for example, 
in LOI of the JPARC-SK experiment in their sensitivity estimate \cite{JPARC}.
Whereas for the green dashed and the blue-dotted lines, the full region 
$0 \leq \theta_{23} \leq \pi/2$ is allowed as output, and the input value 
of $\theta_{23}$ is taken in the first (second) octant of $\theta_{23}$ 
for the green dashed (blue dotted) lines.
We recognize that our ignorance of which octant $\theta_{23}$ lives 
makes the error of $\sin^2{2\theta_{23}}$ almost twice 
($\sim$ 80\% increase) 
in the optimistic (pessimistic) case of systematic errors.

\begin{figure}[tbp]
\begin{center}
\includegraphics[width=0.6\textwidth]{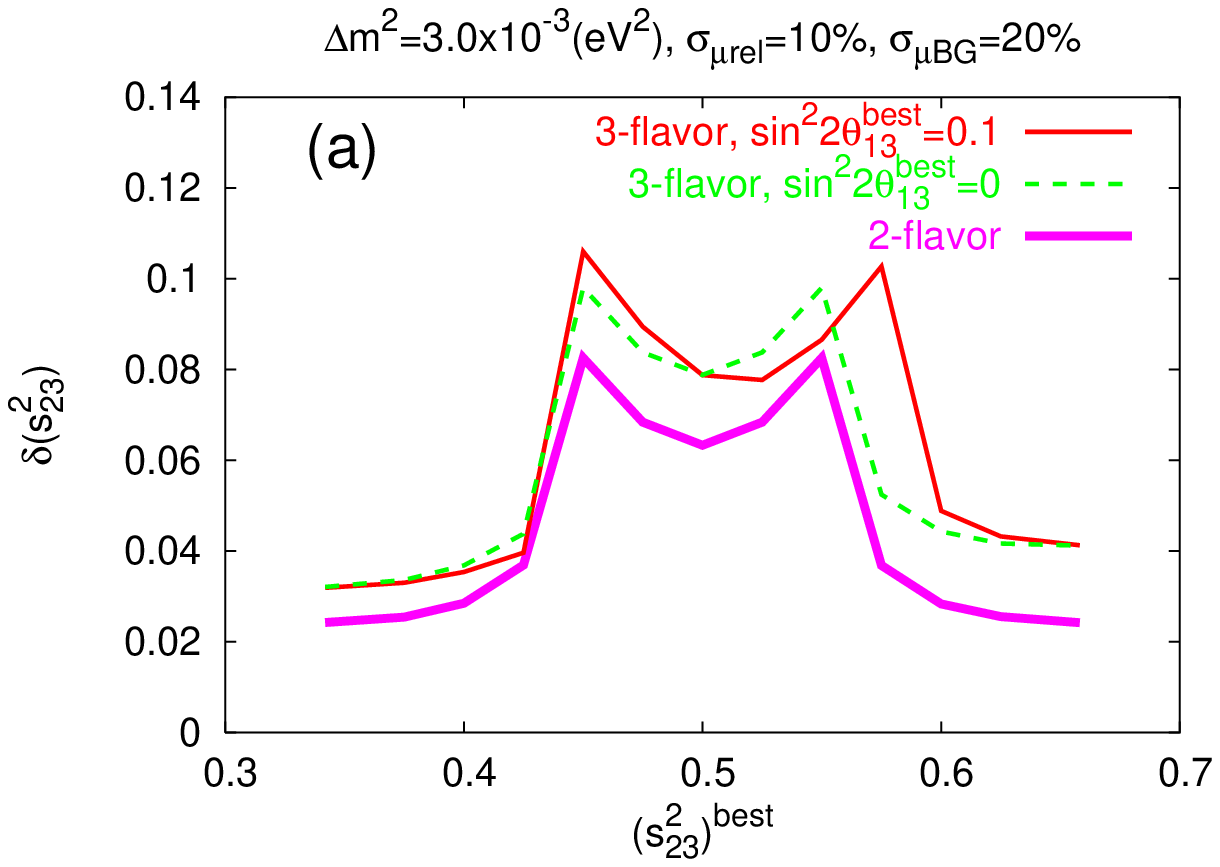}
\includegraphics[width=0.6\textwidth]{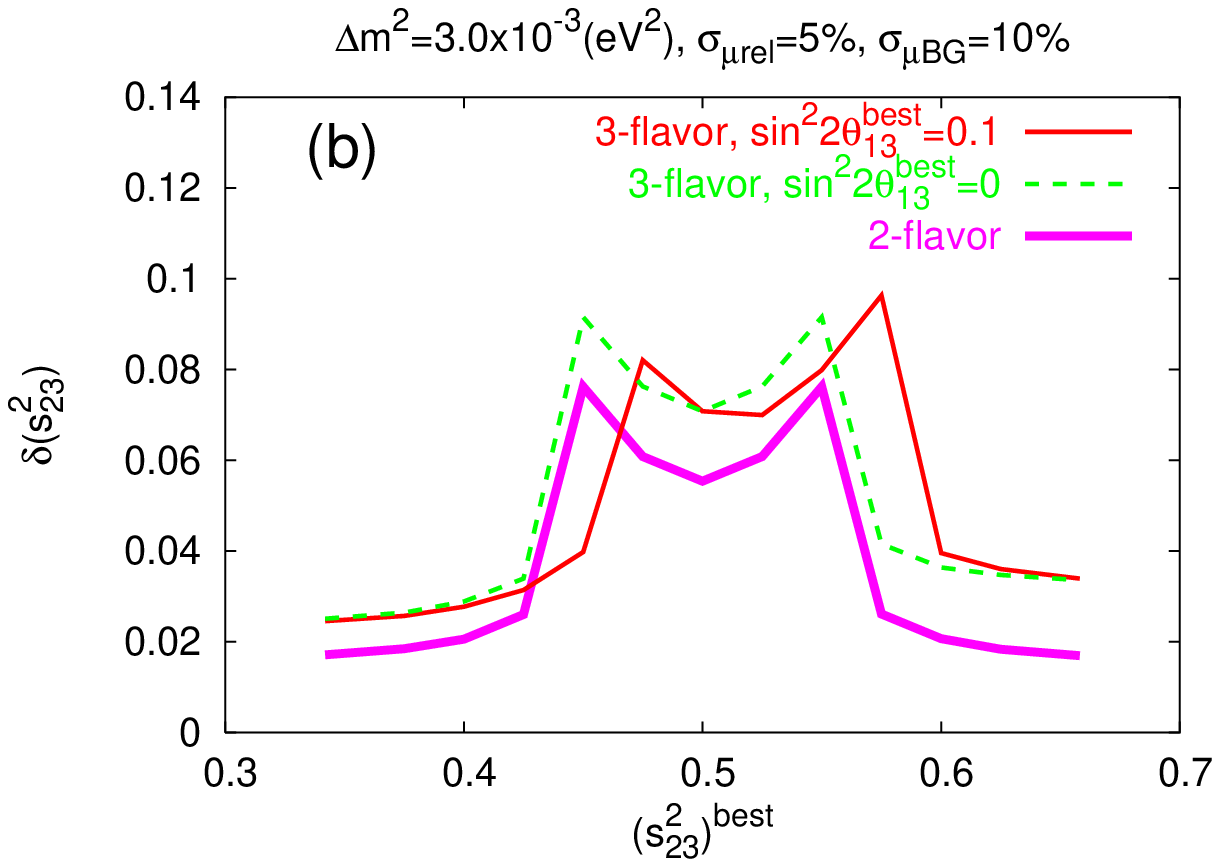}
\end{center}
\caption{
The expected accuracies of determination of 
$s_{23}^2$ at 90\% CL obtained by the three-flavor analysis 
are compared with that of the two-flavor analysis
($\theta_{13}=\theta_{13}^\best =0$)
which is indicated by pink thick-solid line.
The red thin-solid and green dashed lines are for input values of 
$\theta_{13}$ as 
$\sin^2{2\theta_{13}^\best}=0.1$ and $\sin^2{2\theta_{13}^\best}=0.0$, 
respectively. 
Fig.~\ref{ser_30_theta}a and \ref{ser_30_theta}b are with systematic errors 
($\sigma_{\mu \rel}$, $\sigma_{\muBG}$) = (10\%, 20\%) and 
(5\%, 10\%), respectively.
}
\label{ser_30_theta}
\end{figure}

In Fig.~\ref{ser_30_theta}, 
the errors expected for determination of $s^2_{23}$ are plotted, 
with systematic errors 
($\sigma_{\mu \rel}$, $\sigma_{\muBG}$) = (10\%, 20\%) and 
(5\%, 10\%) in Fig.~\ref{ser_30_theta}a and Fig.~\ref{ser_30_theta}b, 
respectively. 
The readers may be surprised by the difference between 
Fig.~\ref{ser_30_2theta} and 
Fig.~\ref{ser_30_theta}. 
Although they are based on exactly the same data set, 
the errors $\delta (s_{23}^2)/s_{23}^2$ in Fig.~\ref{ser_30_theta}
are as large as 14\%-20\% in the ``near to the maximal'' region 
and the shape is quite odd, which should be compared to the 
smooth behavior of the curves 
in Fig.~\ref{ser_30_2theta}. 
Hereafter,  we use the notations 
``near to the maximal'' and ``off the maximal'' regions which mean 
the regions of $\theta_{23}$, 
$|s^2_{23} - 0.5| \leq 0.08$ and 
$0.08 \leq |s^2_{23} - 0.5| \leq 0.16$, respectively. 
There are two main reasons for such a marked difference.

\vskip 0.1cm

\noindent
(a) Jacobian effect

As explained in Sec.~\ref{Jacobian} the Jacobian (\ref{jacobian}) 
makes the uncertainty of $s^2_{23}$ much larger than that of 
$\sin^2{2\theta_{23}}$ near the maximal mixing.
It explains the feature that the uncertainty is large in region 
$0.45 \leq s^2_{23} \leq 0.55$, but it does not explain 
the double-peaked structure in Fig.~\ref{ser_30_theta}. 

\vskip 0.1cm

\noindent
(b) Degeneracy effect

The problem of degeneracy related with $\theta_{23}$ 
requires a special comment. As one can see in Fig.~\ref{th-a_reg}, 
there is a clone solution for a given input value of $\theta_{23}$. 
We do not include the separated clone region in defining the 
uncertainties in $s^2_{23}$. 
We think it appropriate not to include the separate region because 
it is due to the problem of parameter degeneracy and is a separate 
issue from the uncertainty attached to experimental $s^2_{23}$ determination. 
Nevertheless, it affects the results in Fig.~\ref{ser_30_theta} 
because the genuine and the clone regions merge near 
the maximal mixing. Once the true and the fake regions merge 
there is no way to separate them. It creates a jump of $\delta (s_{23}^2)$
at $s^2_{23} \simeq 0.45$ and $\simeq0.55$ when the true value of 
$s^2_{23}$ approaches to 0.5 from above and from below, respectively, 
leading to the double-peaked structure in Fig.~\ref{ser_30_theta}.
\footnote{
Strictly speaking, it may not to be appropriate to connect the points
of $\delta (s_{23}^2)$ before and after the merge in Fig.~\ref{ser_30_theta}
because the rise of $\delta (s_{23}^2)$ is a discrete jump.
}

Notice that the sensitivity curves in Fig.~\ref{ser_30_2theta} obtained not 
only with but also without the restriction $0 \leq \theta_{23} \leq \pi/4$ 
do not have the similar jump.
It is because in the analysis of error defined in terms of the variable 
$\sin^2{2\theta_{23}}$ the true and the fake regions start to 
overlap at even larger deviation from the maximal value of 
$\theta_{23}$, leading to a smooth behavior.
It stems from the fact that the variable $\sin^2{2\theta_{23}}$ cannot 
distinguish between the first and the second octants, through which 
the confusion due to $\theta_{23}$ ambiguity is maximally enhanced. 
It also implies that use of the variable $\sin^2{2\theta_{23}}$ 
in defining errors leads to larger errors in regions where 
$\delta(s^2_{23})$ jumps down in Fig.~\ref{ser_30_theta}.
We emphasize again that one has to use the variable $s^2_{23}$ 
to represent uncertainty in $\theta_{23}$ determination because, 
in addition to the one mentioned just above, 
the angle is defined in the entire quadrant and the obtained 
uncertainties are different in the first and in the second octants.

From Fig.~\ref{ser_30_theta} we also observe the following features. 

\vskip 0.1cm

\noindent
(i) The influence of the systematic error on the uncertainty in 
$s^2_{23}$ determination is rather mild. 
With a factor of 2 smaller systematic errors, 
$\delta (s_{23}^2)$ decreases only by $\simeq$15\% in the near 
to the maximal region and by $\simeq$20\% in off the maximal region.

\vskip 0.1cm

\noindent
(ii) The three-flavor (nonzero $\theta_{13}$) effect exists 
in the entire range of $s^2_{23}$ within the SK allowed region.
Its magnitude differs depending upon which region of $s^2_{23}$ 
and which octants of $\theta_{23}$ is talked about.
Roughly speaking, the three-flavor effect makes $\delta (s_{23}^2)$ increase
by $\simeq 0.015$ (20\%-25\%)
 in the near to the maximal region
and $\simeq 0.008$ ($\simeq 0.017$) in off the maximal region
in the first (second) octant of $\theta_{23}$. 
Note that the uncertainty is almost doubled in the second octant.

\vskip 0.1cm

\noindent
(iii) 
It is indicated in Fig.~\ref{ser_30_theta} that 
the shape of uncertainty curve is distorted in the case of input value 
$\sin^2{2\theta_{13}^\best}=0.1$; 
The ``Mexican hat'' moves slightly to rightward, and 
curiously, better sensitivities are obtained in some regions. 
Peak in the uncertainty curves signals that merging of the degenerate 
solutions occurs, and its exact location is determined by a 
delicate balance between the parameters. Therefore, 
the values of uncertainties around the peak region cannot be trusted 
at a face value. 
However, we have checked that the uncertainty
at $(s^2_{23})^\best=0.5$ does not 
suffer from the degeneracy problem in a severe way, 
as is shown in Fig.~\ref{ser_30_2theta}. 
Hence, we have restricted ourselves into a roughly estimated 
uncertainties in the above discussions. 
The uncertainties are not so sensitive to the input value of $\theta_{13}$ 
apart from this sensitivity to merging degenerate solutions.
 Especially, the uncertainty for $(s^2_{23})^\best=0.5$ is
almost independent of $(s_{13}^2)^\best$
as we can understand with (\ref{PvacOM}).

\vskip 0.1cm

We have also examined the case of 
$\Delta m^2_{31} = 2.5 \times 10^{-3} \mbox{eV}^2$ to 
make comparison with that of 
$\Delta m^2_{31} = 3 \times 10^{-3} \mbox{eV}^2$.
It may worth to examine because it may mimic the situation 
that the JPARC-SK experiment run at slightly higher energies than 
the vacuum oscillation maximum. 
While the qualitative features noted above also hold in this case 
there are some notable changes:

\vskip 0.1cm

\noindent
(iv) With $\Delta m^2_{31} = 2.5 \times 10^{-3} \mbox{eV}^2$, 
the uncertainty in $s^2_{23}$ determination becomes larger 
by $\simeq$15\% in the near to the maximal region and by 
$\simeq$20\% in the off the maximal region.
The feature is essentially the same for both the optimistic 
and the pessimistic systematic errors, 
($\sigma_{\mu \rel}$, $\sigma_{\muBG}$) = (5\%, 10\%) 
and (10\%, 20\%).

\section{To what extent does LBL $\nu_{e}$ appearance or reactor measurement help?}
\label{app+react}

\begin{figure}[tbp]
\begin{center}
\includegraphics[width=0.6\textwidth]{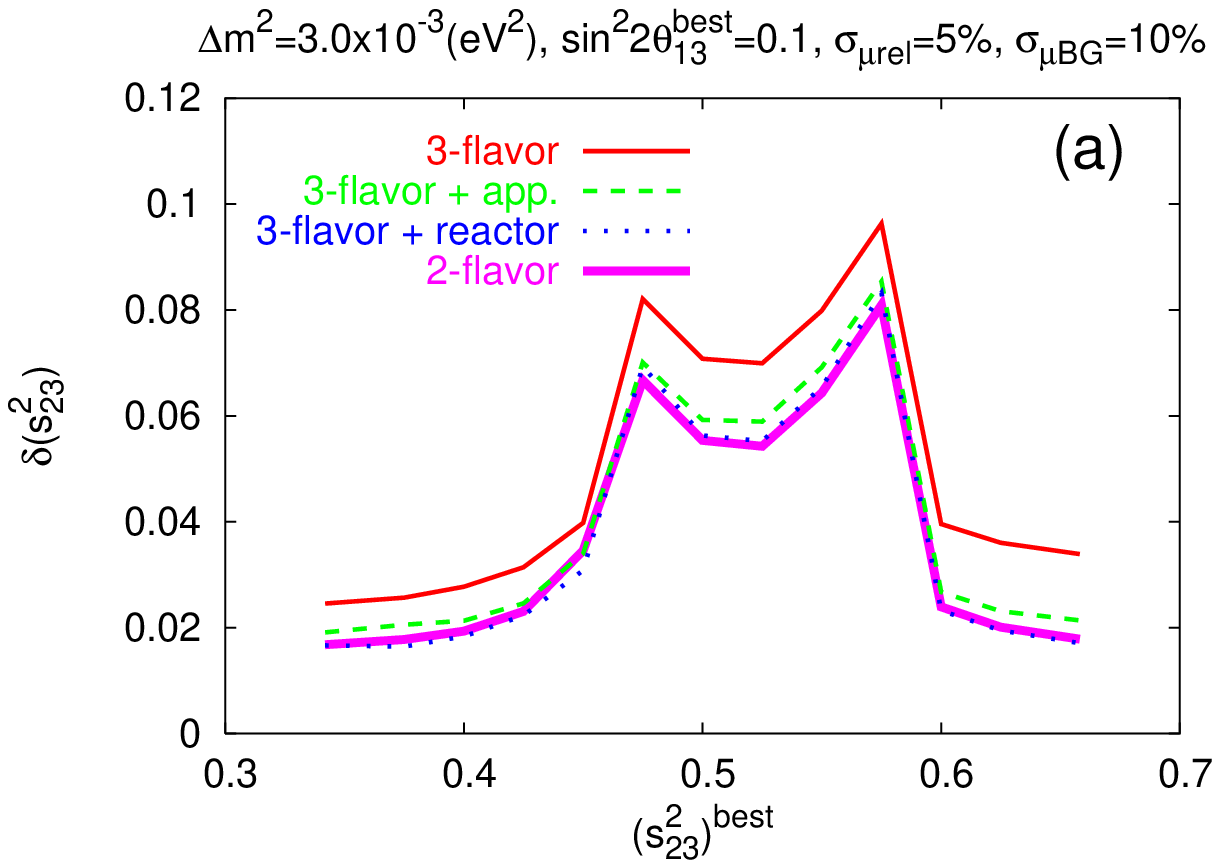}\\
\includegraphics[width=0.6\textwidth]{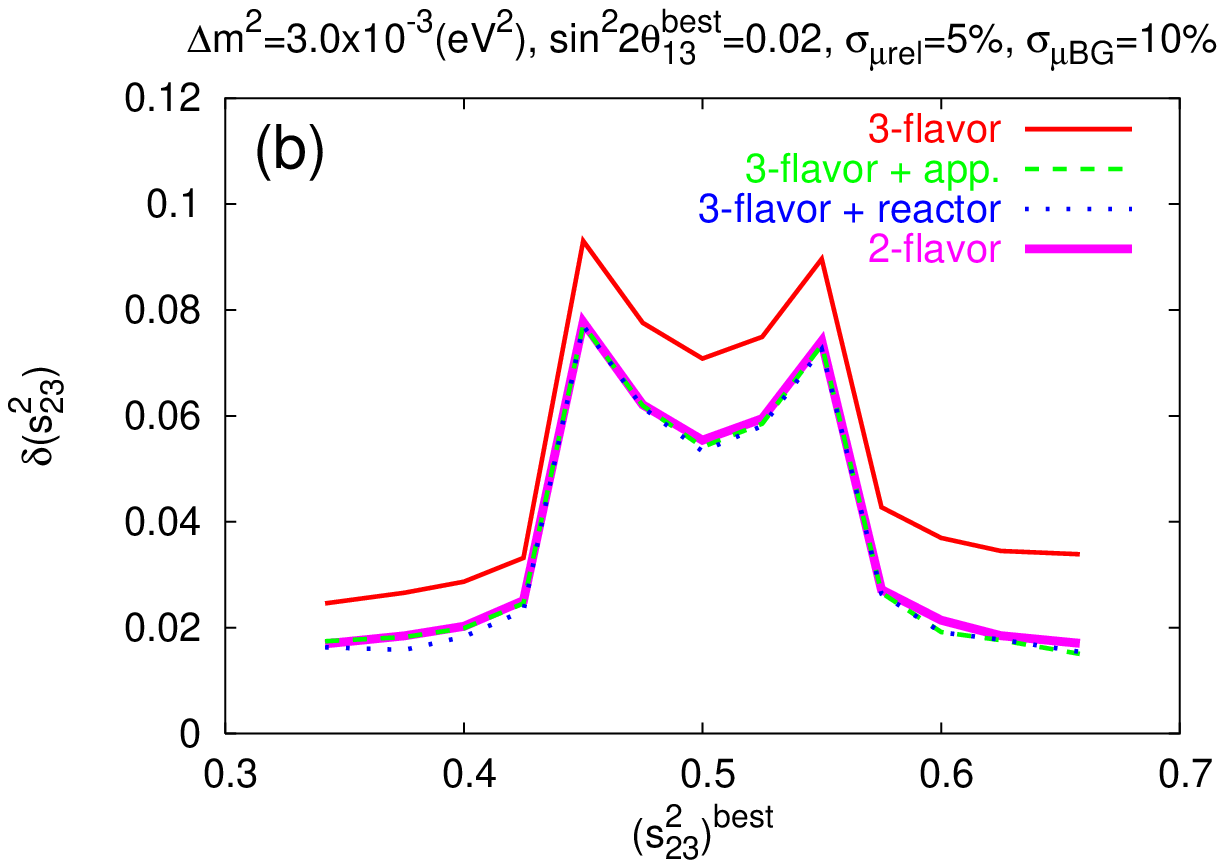}
\end{center}
\caption{
The expected accuracies of $s^2_{23}$ determination at 90\%~CL 
when combined with $\nu_e$ appearance (green dashed line) 
or reactor measurement of $\theta_{13}$ (blue dotted line).
They indicate the improvement over the one reachable by three-flavor 
analysis of $\nu_{\mu}$ disappearance measurement only, assuming 
$\sin^2{2\theta_{13}} \leq 0.15$, which is indicated by red thin-solid line.
The true value of $\sin^2{2\theta_{13}}$ is assumed to be
0.1 and 0.02 in Fig.~\ref{app}a and \ref{app}b, respectively.
The result of two-flavor analysis (pink thick-solid line), which is 
given for comparison, is almost degenerate with green dashed 
and blue dotted lines.
}
\label{app}
\end{figure}

A better sensitivity for $\theta_{23}$ can be achieved by combining 
some other experiments that are sensitive to $\theta_{13}$ with 
disappearance measurement, 
though it cannot solve the problems of the large Jacobian and the 
degeneracy; An improvement can be expected to the extent that the loss of
sensitivity of $\theta_{23}$ is caused by the uncertainty of $\theta_{13}$. 
There are two possibilities for the supplementary experiments: 
$\nu_e$ appearance measurement in LBL experiments and 
$\theta_{13}$ measurement in reactor experiments. 
Fortunately, the $\nu_{\mu}$ disappearance measurement 
is always accompanied by the simultaneous $\nu_e$ appearance 
measurement of $P(\nu_{\mu} \rightarrow \nu_{e})$ in most of 
the LBL experiments. 
Therefore, it is natural to think about combining them first to 
determine $\theta_{23}$ more precisely.
But if a reactor measurement of $\theta_{13}$ is carried out it, 
it may give a better determination of $\theta_{13}$ than 
LBL $\nu_e$ appearance experiments because the latter suffers 
from the additional uncertainty due to $\delta$ \cite{KMN02}.


Here, we present only the result for $\delta^\best = 0$.  
We have checked that when the input value 
$\delta^\best$ is varied in the range 
$- \pi/2 \leq \delta^\best \leq \pi/2$ 
the variation of $\delta (s^2_{23})$ is within $\pm 5\%$
relatively to the value for $\delta^\best = 0$.
Note that the $\Delta \chi^2$ must be minimized with respect 
to $\delta$ because we can not know the value of $\delta$ from
$\nu_e$ appearance only.
Again, we postpone the detailed description of our statistical procedure 
in Appendix C.

The obtained results of how the sensitivity improves with 
LBL $\nu_e$ appearance and reactor measurement are presented 
in Fig.~\ref{app} for two values of $\theta_{13}^\best$, 
$\sin^2{2\theta_{13}^\best}=0.1$ (Fig.~\ref{app}a) 
and 0.02 (Fig.~\ref{app}b).
We examine the cases with two true values of $\sin^2{2\theta_{13}}$
because the accuracy of $\sin^2{2\theta_{13}}$ measurement
in appearance experiments depends on its true value. 
On the other hand, the error of $\sin^2{2\theta_{13}}$ 
in the reactor-neutrino disappearance measurement is 
nearly independent of the true value of $\theta_{13}$,  
$\delta \left(\sin^2{2\theta_{13}} \right) \simeq 0.02$ at 90\%~CL, 
as was shown in our previous analyses~\cite{MSYIS,reactorCP}.
We utilize this feature to simplify our formulation of the 
statistical procedure with reactor measurement in Appendix C.

It is shown in Fig.~\ref{app} that the accuracy of $\theta_{23}$ measurement
is indeed improved thanks to the help by $\nu_e$ appearance experiment 
or reactor measurement of $\theta_{13}$.
The reason why the better sensitivity of $\theta_{23}$ is obtained 
at smaller $\sin^2{2\theta_{13}^\best}$ in the case with appearance measurement 
is that the error of $\sin^2{2\theta_{13}}$ determination increases 
for greater values of $\sin^2{2\theta_{13}^\best}$
due to larger $\sigma_{e \rel}^2 (N^\best_e)^2$ in (\ref{chi2e}).
In fact, the case presented in Fig.~\ref{app}b with 
$\sin^2{2\theta_{13}^\best}=0.02$ represents the best sensitivity 
attainable by combined appearance-disappearance measurement.
Notice that the error of $\theta_{23}$ cannot be smaller than that 
of two-flavor fitting ($\theta_{13} = 0$) because it is the limit of 
no uncertainty in $\theta_{13}$. Moreover, appearance measurement 
itself is not very sensitive to $\theta_{23}$ and hence it plays only a 
supplementary role by constraining $\theta_{13}$.

It is also shown in Fig.~\ref{app} that with reactor measurement of 
$\theta_{13}$ the error of $\theta_{23}$ determination is reduced 
to the one expected in the two-flavor fitting with $\theta_{13}=0$ 
independent of the values of $\theta_{13}^\best$.
Thus, the reactor measurement of $\theta_{13}$ can do a better job 
in improving accuracy of $\theta_{23}$ determination in relatively 
large $\theta_{13}^\best$.  
It is due to the feature that the uncertainty in $\theta_{13}$ 
determination in reactor experiments is insensitive to the 
true values of $\theta_{13}$, and that it is a pure measurement 
of $\theta_{13}$. 
In the combined analysis with appearance measurement, by contrast,
$\min_\delta \Delta \chi_e^2$ (see Appendix C for definition) 
vanishes not only at 
the best fit point but also in region around it 
due to the minimization in terms of $\delta$, and 
no improvement of $\Delta \chi_\mu^2$ is obtained in the region.

We did not attempt to further combine $\nu_{e}$ appearance 
measurement, because the sensitivity in two-flavor analysis, 
which can be regarded as the limiting sensitivity attainable 
by the present method, is already achieved with  
reactor measurement of $\theta_{13}$.

\section{Concluding remarks}

In the next generation LBL neutrino oscillation experiments, a
precise determination of $\sin^2{2\theta_{23}}$ to a few \% level
is expected.
In this paper, we have pointed out that, 
despite such marvelous sensitivities for $\sin^2{2\theta_{23}}$, 
the reachable accuracy of determination of $s^2_{23}$ is only at 
a level of 
$\delta (s_{23}^2) / s_{23}^2 = 10\%\mbox{-}20\%$ 
depending upon the true value of $\theta_{23}$.

We have identified the principal reasons for the disparity between 
the errors of $\sin^2{2\theta_{23}}$ and $s^2_{23}$ as due to the 
Jacobian and the degeneracy effects. 
We also noticed that 
measurement of $\theta_{23}$, when it goes down to such high
accuracy, is started to be affected by the higher order corrections 
in $s_{13}^2$ in the oscillation probability.  
We have shown that by doing detailed quantitative analyses 
such three-flavor effects produce, typically, an increase
of uncertainty in $s^2_{23}$ determination
by $0.008 \mbox{-} 0.017$;
$\delta(s^2_{23})$ is larger in the second octant of $\theta_{23}$ 
than that in the first octant.


Then, we have shown that by combining either reactor measurement of
$\theta_{13}$, or $\nu_e$ appearance measurement in LBL experiment
the accuracies of $\theta_{23}$ determination can be improved to
the level expected by simple two-flavor fitting ($\theta_{13} = 0$)
of the measurement, the limiting sensitivity attainable by the 
disappearance measurement.\footnote{
In fact, accelerator $\theta_{13}$ measurement is affected
by the disappearance experiment through $s_{23}^2$ and/or 
$\Delta m_{31}^2$. Therefore, we may need
simultaneous determination of $\theta_{23}$ and $\theta_{13}$
for the most stringent bounds. 
While we did not go into deep into the detail on how such
analysis for simultaneous determination of two angles can be done, 
such a proper formulation is necessary and should be ready 
when the JPARC-SK experiment starts.
}

We have to note that our simulation of data and the
estimation of uncertainties were done with OA 2 degree beam, which
may be slightly higher in peak energy than the one appropriate
for $\Delta m^2_{31} = 2 \mbox{-} 2.5 \times 10^{-3}$eV$^2$.
It is worthwhile to repeat the analysis with more realistic
OA 2.5 degree beam.

Finally, we want to make some remarks on importance of precise 
determination of the lepton flavor mixing parameters.  
We want to stress two viewpoints, one related to physics and the 
other to future neutrino experiments. 

\vskip 0.1cm

\noindent
(1) Why is precise measurement of the lepton flavor mixing parameters
so important?


One of the significant features is that there are two large and small mixing 
angles. It is suggested that near maximal $\theta_{23}$ may imply a 
symmetry and $\theta_{23}$ is maximal and $\theta_{13}$ vanishes 
in the symmetry limit~\cite{symmetry}. 
If it is indeed the case detecting small deviation of 
$\theta_{23}$ from the maximal is as important as measuring 
$\theta_{13}$. Confirmation of this possibility would 
require a few~\% level determination of $s^2_{23}$, 
given already severe constraint 
by the global analysis, $s^2_{13} \leq 0.047$ at 3$\sigma$~CL \cite{global}.
If it turned out that $\theta_{13}$ is much smaller, it would necessitate 
even more accurate determination of $\theta_{23}$.
If some of the lepton and quark mixing angles 
are indeed ``complementary'' with each other~\cite{mina-smi}, 
there will be immense requests for precise determination of 
lepton mixing angles because the quark mixing angles are now 
measured quite accurately. 
This possibility may require even more precise determination of 
$\theta_{23}$ beyond a level expectable by current technology.

Keeping in mind the possibility that we finally uncover the 
real theory of flavor mixing in the future it is desirable, 
whenever opportunity exists, to pursue determination of 
mixing parameters as precisely as possible. 
Notice that there is a way to measure $\theta_{12}$ to an accuracy of 
$\simeq$ 2\% comparable to that of the Cabibbo angle \cite{MNTZ}.

\vskip 0.1cm

\noindent
(2) How important or crucial is the precise determination of 
$\theta_{23}$ for future neutrino oscillation experiments?

Most probably the field where precise determination of 
lepton mixing parameters is really necessary is the detection of 
leptonic CP violation, one of the ultimate goal of future 
neutrino experiments. 
It is because CP violation is a tiny effect, being suppressed by the 
two small numbers, one the Jarlskog factor 
$\simeq 0.04 \sin{\delta}$ and the other 
$\epsilon \equiv \Delta m^2_{21} / \Delta m^2_{31} \simeq 0.033$.
Hence, it is crucial to minimize uncertainties of other mixing parameters 
(as well as neutrino cross sections on the relevant nuclei) 
to achieve unambiguous detection of CP violation.

In this regard, diminishing the uncertainty in $\theta_{23}$ measurement is 
probably one of the most important. To obtain the feeling let us 
suppose for simplicity that the experiment has relatively short 
baseline so that the vacuum oscillation approximation applies, 
which may be the case for  the JPARC-Hyper-Kamiokande project
\cite{Hyper-K}. Let us assume that the CP asymmetry defined by  
$A \equiv (P(\nu)-P(\bar{\nu}))/(P(\nu)+P(\bar{\nu}))$ 
gives a good measure for CP violation, where 
$P(\nu)$ and $P(\bar{\nu})$ denote appearance probabilities 
$P(\nu_{\mu} \rightarrow \nu_{e})$ and 
$P(\bar{\nu}_{\mu} \rightarrow \bar{\nu}_{e})$, respectively.
To simplify the equations we tune the energy at the first
oscillation maximum,
$\Delta m^2_{31} L / 2 E = \pi$,
as planned in the J-PARC experiment.
Under these approximations, the CP asymmetry is given by
\begin{eqnarray}
A \equiv
\frac{P(\nu)-P(\bar{\nu})}{P(\nu)+P(\bar{\nu})} = -
\frac{\sin{2 \theta_{12}}\cot{\theta_{23}}}{s_{13}}
\left(\frac{\Delta m^2_{21} L}{4 E}\right) \sin\delta .
\label{cpasym}
\end{eqnarray}

At the moment, of course, the uncertainty of $s_{13}$ is 
a dominating error because it has not been measured yet. 
But it is expected that it can be measured either by LBL or 
reactor experiments with errors $\simeq 10 \mbox{-} 25$\% if 
$\sin^2{2\theta_{13}} > 0.04$. 
The error produced by $\cot{\theta_{23}}$ factor in (\ref{cpasym}) 
is non-negligible compared to this error. 
The uncertainties estimated in Sec.~\ref{analysis} can be translated into 
$\simeq 5 \mbox{-} 15$\% error in $\cot{\theta_{23}}$, and hence it is large.
Here, it is worth to note that where the errors are relatively small the 
splitting of $\cot{\theta_{23}}$ between two degenerate solutions of 
$\theta_{23}$ is large. 
Therefore, the uncertainty in the CP asymmetry is even larger, 20-30\%,  
in off the maximal region if the octant degeneracy of $\theta_{23}$ is not resolved. 
Most likely, the error of $s_{13}$ is significantly reduced at the 
time of CP measurement because of the possibility that it 
is precisely determined by {\it in situ} LBL measurement 
employing both neutrino and antineutrino beams at the oscillation 
maximum (or more precisely the shrunk ellipse limit~\cite{KMN02}).
Therefore, the uncertainty of $\theta_{23}$ including the degeneracy effect 
may be the dominating error for CP measurement.

\vskip 0.1cm

Precise determination of $\Delta m^2_{31}$, though not explicitly 
addressed in this paper, is an another important issue. 
The error is expected to be $\sim 10$\% in the JPARC-SK experiment. 
Its precise determination is also essential in a variety of contexts ranging 
search for CP violation and improving 
bounds on mixing parameters by double beta decay experiments.

In conclusion, we stress the importance of finding alternative methods 
which are sensitive to $s_{23}$ not to $\sin{2 \theta_{23}}$ 
for accurate measurement of $\theta_{23}$. 
The obvious candidates include the reactor-LBL combined measurement 
discussed in \cite{MSYIS}, and high-statistics observation of atmospheric 
neutrinos \cite{kajita_noon04,concha-smi_23}.

\newpage

\appendix
\section{Derivation of disappearance probability 
$P(\nu_{\mu} \rightarrow \nu_{\mu})$ 
to the next to leading order in $\Delta m^2_{21} / \Delta m^2_{31}$ 
and $s^2_{13}$}

In this Appendix, we give a self-contained discussion for 
deriving the expression of disappearance probability 
$P(\nu_{\mu} \rightarrow \nu_{\mu})$
which is valid to the next to leading order in $\epsilon$ 
and $s^2_{13}$. We use the method developed by 
Kimura, Takamura, and Yokomakura (KTY)~\cite{KTY}.

The evolution equation of neutrinos can be written in the 
flavor eigenstate as 
\begin{eqnarray}
i\frac{d}{dx} \nu_{\alpha} = 
\frac{1}{2E} H_{\alpha \beta} \nu_{\beta}
\hspace*{15mm}
(\alpha, \beta = e, \mu, \tau),
\label{evolution}
\end{eqnarray}
where the Hamiltonian is given by
\begin{eqnarray}
H=
U \left[
\begin{array}{ccc}
\Delta m^2_{11} & 0 & 0 \\
0 & \Delta m^2_{21}& 0 \\
0 & 0 & \Delta m^2_{31} 
\end{array}
\right] U^{+}
+
\left[
\begin{array}{ccc}
a(x) & 0 & 0 \\
0 & 0 & 0 \\
0 & 0 & 0
\end{array}
\right], 
\label{hamiltonian}
\end{eqnarray}
whose first term will be denoted as $H^{vac}$ hereafter 
and $\Delta m^2_{ji} \equiv m^2_{j} - m^2_{i}$. 
(Hence, $\Delta m^2_{11} \equiv 0$ by definition.)
In (\ref{hamiltonian}), $a \equiv 2\sqrt{2} G_F N_e(x) E$ 
denotes the index of refraction of neutrinos in medium of electron 
number density $N_e(x)$, where $G_F$ is the Fermi constant and 
$E$ is the neutrino energy~\cite{wolfenstein}.
Despite that $N_e(x)$ may depend upon locations along 
the neutrino trajectory, we use constant density 
approximation throughout this paper. 
The MNS matrix $U$ relates the flavor and the vacuum 
mass eigenstates as 
\begin{eqnarray}
\nu_{\alpha} = (U)_{\alpha i} \nu_{i} ,
\end{eqnarray}
where $i$ runs over 1-3.

We now define the mass eigenstate in matter $\nu_{i}^{m}$ 
by using transformation 
\begin{eqnarray}
\nu_{\alpha} = (V)_{\alpha i} \nu_{i}^{m} ,
\end{eqnarray}
where $V$ is the unitary matrix which diagonalize the Hamiltonian 
with scaled eigenvalues $\lambda$ as 
$V^{\dagger} H V = H_{diag} 
\equiv diag(a\lambda_1, a\lambda_2, a\lambda_3)$. 
We first obtain the expressions of the eigenvalues of the 
Hamiltonian (\ref{hamiltonian}). 
They are determined by the equation 
$det[H - \lambda a I]=0$ which is the cubic equation 
for the scaled eigenvalue $\lambda$:
\begin{eqnarray}
\lambda^3 - 
(\delta_{21} + \delta_{31} +1) \lambda^2 +
\left[
(c_{12}^2 c_{13}^2 + s_{13}^2) \delta_{21} + 
c_{13}^2 \delta_{31} + \delta_{21} \delta_{31}
\right] \lambda - 
c_{12}^2 c_{13}^2 \delta_{21} \delta_{31} = 0 . 
\end{eqnarray}
Notice that everything is scaled by $a$ and 
$\delta_{21}$ and $\delta_{31}$ denote the scaled squared 
mass differences,
\begin{eqnarray}
\delta_{21} \equiv \frac{\Delta m^2_{21}}{a}, 
\hspace*{15mm}
\delta_{31} \equiv \frac{\Delta m^2_{31}}{a}.
\end{eqnarray}
To first order in $\delta_{21}$ and $s^2_{13}$ 
the solutions of the equation are given under the convention 
that $\lambda_1 < \lambda_2 < \lambda_3$ by 
\begin{eqnarray}
\lambda_1 &=& c_{12}^2 \delta_{21} ,
\nonumber \\
\lambda_2 &=& 1 + s_{12}^2 \delta_{21} - 
\frac{\delta_{31}}{\delta_{31}-1} s_{13}^2 ,
\nonumber \\
\lambda_3 &=& \delta_{31} + 
\frac{\delta_{31}}{\delta_{31}-1} s_{13}^2 .
\label{eigenvalue}
\end{eqnarray}
Notice that we are in the intermediate energy region between 
high-energy (atmospheric) and low-energy (solar) level crossings. 
If we sit on the high-energy region above two level crossings, 
the expressions for $\lambda_2$ and $\lambda_3$ must be 
interchanged. 
We have checked that the same analytic formulas are obtained, 
when expressed in terms of observable physical quantities,   
even if we work in the high-energy region.

We follow the KTY method for deriving 
$P(\nu_{\mu} \rightarrow \nu_{\mu})$ and write down the equations 
\begin{eqnarray}
H_{\mu \mu} &=& H^{vac}_{\mu \mu}, 
\nonumber \\
H_{e e} H_{\tau \tau} - H_{e \tau} H_{\tau e} &=&
(H^{vac}_{e e} + a) H^{vac}_{\tau \tau} - 
H^{vac}_{e \tau} H^{vac}_{\tau e}. 
\label{KTYeq1}
\end{eqnarray}
They give relationships between mixing matrix in vacuum and in matter 
as 
\begin{eqnarray}
\sum_{i} \lambda_{i} |V_{\mu i}|^2 &=&
\sum_{i} \frac{\Delta m^2_{i1}}{a} |U_{\mu i}|^2
\equiv p, 
\nonumber \\
\sum_{ijk}^{cyclic} \lambda_{i}\lambda_{j} |V_{\mu k}|^2 &=& 
\sum_{ijk}^{cyclic} \frac{\Delta m^2_{i1} \Delta m^2_{j1}}{a^2}|U_{\mu k}|^2 
+ \sum_{i=1}^{3} \frac{\Delta m^2_{i1}}{a}|U_{\tau i}|^2
\equiv q.
\label{KTYeq2}
\end{eqnarray}
Solving (\ref{KTYeq2}) for $|V_{\mu i}|^2$ 
under the constraint of unitarity 
\begin{eqnarray}
\sum_{i} |V_{\mu i}|^2 = 1 
\end{eqnarray}
we obtain 
\begin{eqnarray}
|V_{\mu i}|^2 = \frac{1}{\Delta_{ji} \Delta_{ki}}
\left[
\lambda_{i} p + q - \lambda_{i} (\lambda_{j}+\lambda_{k})
\right] ,
\end{eqnarray}
where $\Delta_{ji} \equiv \lambda_{j} - \lambda_{i}$ and 
($i,j,k$) are cyclic.

The disappearance probability $P(\nu_{\mu} \rightarrow \nu_{\mu})$ 
is given by 
\begin{eqnarray}
P(\nu_{\mu} \rightarrow \nu_{\mu}) = 1 - 4 \sum_{(i,j)}^{cyclic}
\frac{L_{ij}}
{\Delta_{ij} \Delta_{12} \Delta_{23} \Delta_{31}}
\sin^2{
\left(\frac{\Delta_{ij}aL}{4E}\right)},
\label{disapp_P}
\end{eqnarray}
where $L_{ij}$ is given by 
\begin{eqnarray}
L_{ij} &=&
p^2 \lambda_{i}\lambda_{j} + q^2 + pq(\lambda_{i} + \lambda_{j}) -
p \lambda_{i}\lambda_{j}(\lambda_{i} + \lambda_{j} + 2\lambda_{k})
\nonumber \\
& & 
- q \Bigl[\lambda_{i}(\lambda_{j} + \lambda_{k}) + 
\lambda_{j}(\lambda_{i} + \lambda_{k}) \Bigr] + 
\lambda_{i}\lambda_{j}(\lambda_{i} + 
\lambda_{k})(\lambda_{j} + \lambda_{k}).
\label{disapp_C}
\end{eqnarray}
Notice that $p$ and $q$ are expressed by the vacuum mixing 
parameters through (\ref{KTYeq2}). 
If we insert the exact form of the eigenvalues $\lambda_{i}$ 
in (\ref{disapp_C}) we obtain the exact expression of the 
disappearance probability $P(\nu_{\mu} \rightarrow \nu_{\mu})$, 
which should be identical with 
$1-P(\nu_{\mu} \rightarrow \nu_{e})-P(\nu_{\mu} \rightarrow \nu_{\tau})$ 
computed by KTY~\cite{KTY}.

In this paper, we restrict ourselves into the perturbative expansion 
and keep terms to order $\delta_{21}$, $s^2_{13}$, and 
$\delta_{21} s_{13}$ for CP $\delta$ sensitive terms.
Using the perturbative result of the eigenvalues 
(\ref{eigenvalue}) we obtain, for example, 
\begin{eqnarray}
\frac{1}{\Delta_{12} \Delta_{23} \Delta_{31}} &=& 
\frac{1}{\delta_{31}(\delta_{31}-1)}
\left[1 - 
\frac{\delta_{21}}{\delta_{31}(\delta_{31}-1)} 
\Bigl(
\delta_{31}^2 (s_{12}^2-c_{12}^2) - 2 \delta_{31} s_{12}^2 + c_{12}^2
\Bigr)
\right.
\nonumber \\
& & \left.
+ \frac{s_{13}^2}{(\delta_{31}-1)^2}
\left(
\delta_{31}^2 - 4\delta_{31} + 1
\right)
\right].
\end{eqnarray}
Collecting these formulas, we obtain the expression of 
the disappearance probability to the next to leading order 
as\footnote{
While we were writing the manuscript of this paper, 
we have informed~\cite{ohlsson} that a similar perturbative 
formula was obtained by Akhmedov {\it et al.}~\cite{perturbative}.
We thank Tommy Ohlsson for kindly informing us the paper prior 
submission to the Archive and for verifying the consistency 
between their and our results.
}
\begin{eqnarray}
1 - P(\nu_{\mu} \rightarrow \nu_{\mu}) =  
4(D_{12} + D_{23} + D_{31}) ,
\label{survivalP}
\end{eqnarray}

\begin{eqnarray}
D_{12} + D_{23} = 
\frac{\delta_{31}^2}{(\delta_{31}-1)^2}
\left[
s_{13}^2s_{23}^2 -  
2 \delta_{21} \frac{(\delta_{31}-1)}{\delta_{31}}
J^{\prime}_{r} \cos{\delta}
\right] 
\nonumber \\
\times
\left[
c_{23}^2 \sin^2\left(\frac{aL}{4E}\right) + 
s_{23}^2 \sin^2\left(\frac{aL}{4E} (\delta_{31}-1) \right) 
\right] ,
\label{D1223}
\end{eqnarray}

\begin{eqnarray}
D_{31} &=& 
s_{23}^2c_{23}^2
\sin^2\left(\frac{aL}{4E}\delta_{31}\right) \nonumber \\
&+& 
s_{23}^2c_{23}^2
\left[
- s_{13}^2 
\frac{\delta_{31}^2}{(\delta_{31}-1)^2}
\sin^2\left(\frac{aL}{4E}\delta_{31}\right)
+ \Bigl(-\delta_{21}c_{12}^2+\frac{\delta_{31}}
{\delta_{31}-1}s_{13}^2\Bigr)\Bigl(
\frac{aL}{4E}\Bigr)
\sin\left(
\frac{aL}{2E}\delta_{31}
\right)
\right] \nonumber \\
&+& 2
\frac{\delta_{21}}{\delta_{31}(\delta_{31}-1)}
\Bigl(s_{23}^2 \delta_{31}^2 + 
c_{23}^2 - s_{23}^2 \Bigr)
J^{\prime}_{r} \cos{\delta} 
\sin^2\left(\frac{aL}{4E}\delta_{31}\right) ,
\label{D31}
\end{eqnarray}
where $J^{\prime}_{r} \equiv c_{12}s_{12}c_{23}s_{23}s_{13}$.
The leading order term appears only in $D_{31}$ (the first term).
Notice that once we employ perturbation expansion by 
$\delta_{21} = \Delta m^2_{21}/a$ we cannot take the vacuum 
oscillation limit $a \rightarrow 0$ in (\ref{D1223}) and (\ref{D31}).

We also note a pathology in doing perturbative computation of 
the oscillation probabilities. 	In energy region around dip 
at the oscillation maxima, the leading term in the survival 
probability vanishes and the perturbative expansion is not 
well defined. In fact, the perturbative formula for 
$P(\nu_{\mu} \rightarrow \nu_{\mu})$ 
in (\ref{survivalP}) becomes slightly negative in the energy region, 
To obtain better approximation we may have to sum up small terms 
that are negligible except for this region.  
In the context of the analysis in this paper, however, 
we have explicitly checked that the pathological feature 
does not affect our estimation of the sensitivity.

\section{Method for statistical analysis of disappearance experiment}

We define our statistical method to quantify our 
sensitivity analysis for determination of $\theta_{23}$.
We use the following standard form of $\Delta \chi^2_{\mu}$:
\begin{eqnarray}
\Delta \chi^2_{\mu}
(s_{23}, \Delta m_{31}^2, s_{13})
\equiv \sum_i \frac{\left( N_{\mu i} - N_{\mu i}^\best \right)^2}
{N^\best_{\mu i} + B_{\mu i} +
(\sigma_{\mu \rel} N^\best_{\mu i})^2 + 
(\sigma_{\muBG} B_{\mu i})^2},
\label{chi2mu}
\end{eqnarray}
where $N_{\mu i}$ denotes the number of 
$\nu_\mu$ charged-current quasi-elastic events of the reaction 
$\nu_\mu + N \rightarrow \mu^{-} + N'$ 
in $i$th energy bin which is to be tested against 
the artificially created ``experimental event number'' 
whose best fit is $N_{\mu i}^\best$. 
$B_{\mu i}$ is the number of background events 
calculated in the similar way.
We use 4 energy bins of width 0.2GeV in the energy range 
$E_\nu = 0.4 \mbox{-} 1.2$GeV\@. 
$\sigma_{\mu \rel}$ and $\sigma_{\muBG}$ represent the 
systematic errors associated with signal and 
background events, respectively.\footnote{
$\sigma_{\mu \rel}$ denotes the relative normalization error
between the numbers at front and far detectors,
which corresponds to $\sqrt{2}$ times
the uncorrelated error (e.g., of flux) between those detectors.
We assume, for simplicity, that the baseline length of 
the front detector is short enough to make the numbers of 
events independent of oscillation parameters
and neglect the background as well as the statistical error 
for the numbers of events in the detector.
}

To calculate the numbers of signal events we convolute the survival 
probability $P(\nu_{\mu} \rightarrow \nu_{\mu})$ with the neutrino flux 
times cross sections given in \cite{kameda}. We use for 
$P(\nu_{\mu} \rightarrow \nu_{\mu})$ the expression valid to next to 
leading order in $\epsilon$ and $s^2_{13}$, 
(\ref{survivalP}) with (\ref{D1223}) and (\ref{D31}). 
We have explicitly checked that $\delta$-dependent terms gives a 
minor effect in our analysis because 
they are of the order of $\epsilon s_{13}$ which are higher order 
than terms we are keeping (assuming $\epsilon$ and $s^2_{13}$ are 
comparable in size). 
We assume 5 years running of neutrino mode $\nu_{\mu}$ disappearance 
measurement.
The estimation of background is done by using energy distribution 
of background events after cut kindly provided by the JPARC-SK 
group~\cite{private}.
We have cross-checked our procedure against the numbers of events 
after cut.

The procedure of calculating the error of $\theta_{23}$ determination is as follows. 
We take an input value of $\theta_{13}$ as experimental best fit value
(or nature's choice),
and obtain 90\%~CL allowed region in
$s^2_{23} \mbox{-} \Delta m^2_{31}$, or 
$\sin^2{2\theta_{23}} \mbox{-} \Delta m^2_{31}$
plane.
To obtain the allowed region, we project three-dimensional 
manifold of $\Delta \chi^2_{\mu} =$ {\it constant} onto the 
above two-dimensional plane by imposing the CHOOZ constraint 
$s^2_{23} < 0.04$ ($\sin^2{2\theta_{13}} < 0.15$) at 90\%~CL\@. 
 We rely on the analysis of 2 degrees of freedom
because of the resulting two-dimensional plane.
Notice that while we take a particular value of $\theta_{13}$ 
as an input a wide unlimited region is allowed for output value of 
$\theta_{13}$, as shown in Fig.~\ref{th-a_reg}. 
The feature arises because the disappearance measurement is poor
at restricting $\theta_{13}$. 
Then, $\delta(\sin^2{2\theta_{23}})$ and $\delta(s_{23}^2)$ 
plotted respectively in Fig.~\ref{ser_30_2theta} and Fig.~\ref{ser_30_theta}
are half width of the projected allowed region 
onto the $\sin^2{2\theta_{23}}$ and $s^2_{23}$ axis for a value of $\Delta m^2_{31}$. 
We examine cases of two input values of $\theta_{13}$, 
$\sin^2{2\theta_{13}^\best}= 0.0$ and 0.1 ($(s^2_{13})^\best = 0.026$).

\begin{figure}[tbp]
\begin{center}
\includegraphics[width=0.6\textwidth]{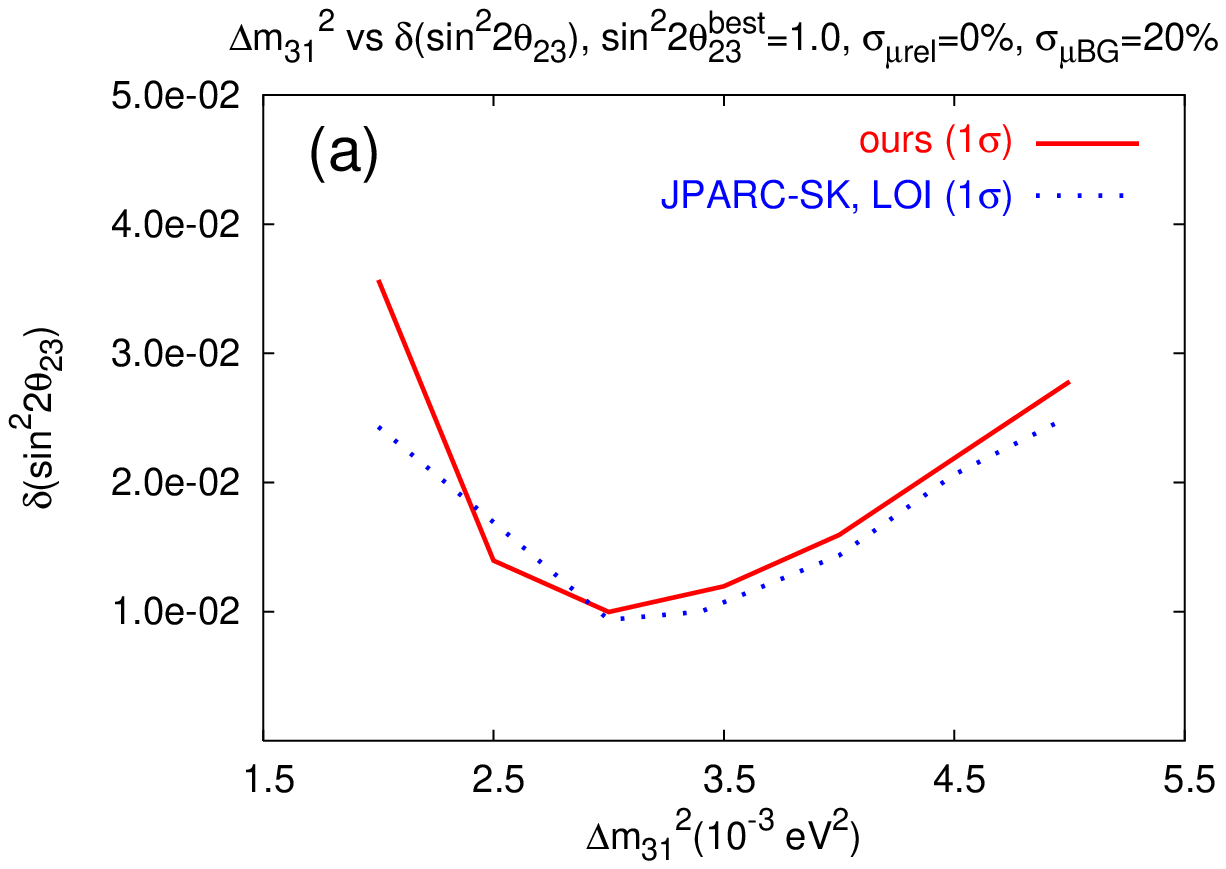}
\includegraphics[width=0.6\textwidth]{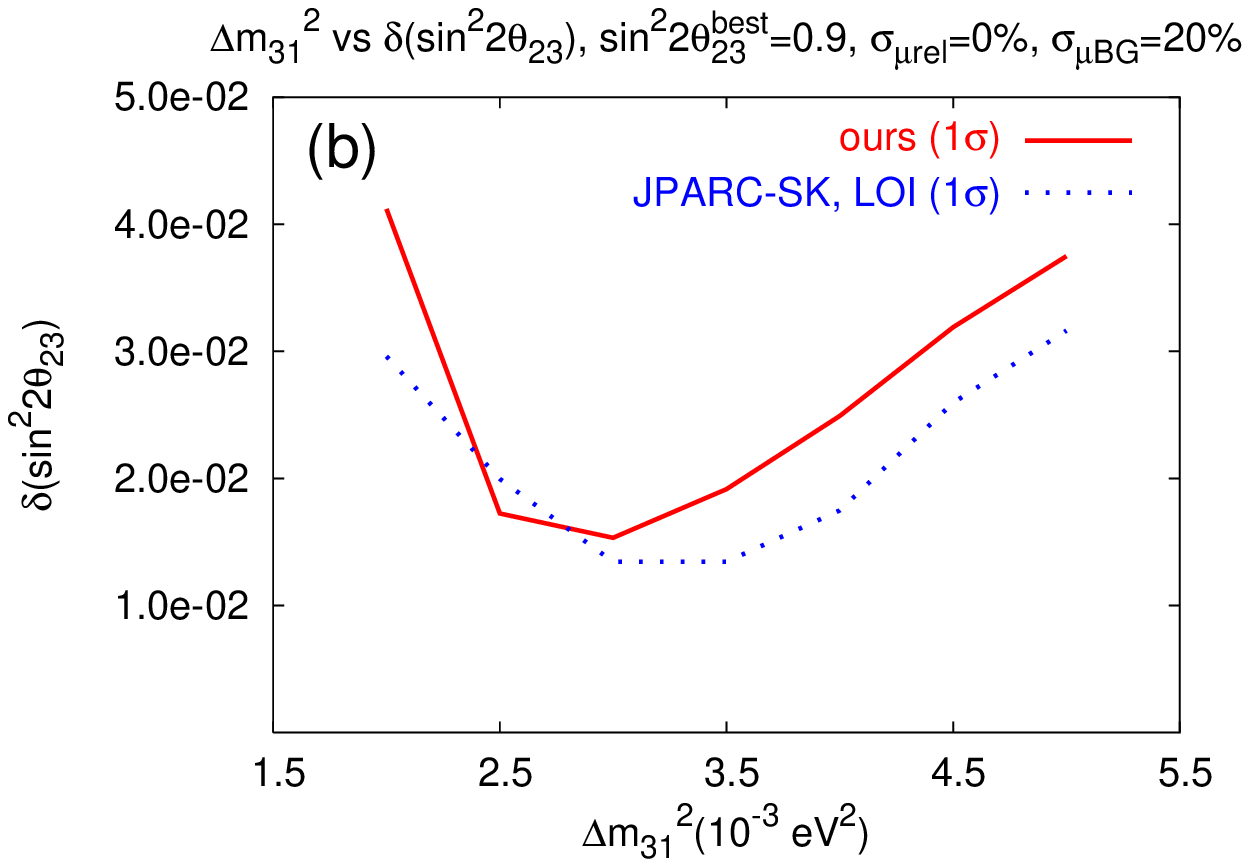}
\end{center}
\caption{Comparison of our estimate of sensitivity 
for $\sin^2{2\theta_{23}}$ (red solid line) 
with that of the JPARC-SK group given in LOI (blue dotted line). 
Both estimates use the two-flavor mixing and 
the same systematic errors. 
The input value of $\theta_{23}$ is taken as maximal, 
$\sin^2{2\theta_{23}^\best}=1$, in Fig.~\ref{LOI_comp}a and as 
$\sin^2{2\theta_{23}^\best}=0.9$ in Fig.~\ref{LOI_comp}b. 
}
\label{LOI_comp}
\end{figure}

In the JPARC-SK's LOI~\cite{JPARC} they describe the procedure 
of determination of $\theta_{23}$ by the $\nu_{\mu}$ disappearance 
measurement at the neutrino energy that corresponds to the first 
oscillation maximum. In Fig.~11 of \cite{JPARC} (original version) 
they present, based on analysis assuming the two-flavor mixing, 
the accuracy of determination of 
$\sin^2{2\theta_{23}}$ and $\Delta m^2_{31}$ as a function of 
$\Delta m^2_{31}$. 
To check the consistency of our procedure of computing event 
numbers as well as the method for statistical analysis 
we try to reproduce the former figure and the result of this 
analysis is given in Fig.~\ref{LOI_comp}; 
Fig.~\ref{LOI_comp}a and 
Fig.~\ref{LOI_comp}b are for the cases 
$\sin^2{2\theta_{23}^\best}=1.0$ and 0.9, respectively, and we have used 
the two-flavor analysis. 
We restrict ourselves to the plot of uncertainty 
of $\sin^2{2\theta_{23}}$ because it is of importance for our 
later analyses.
We use $\sigma_{\mu \rel}=0\%$ and 
$\sigma_{\muBG}=20\%$, 
and take the earth matter density as
$\rho = 2.8\,\mbox{g}\cdot\mbox{cm}^{-3}$ as in LOI\@.\footnote{
Despite the impression one has from the description in the LOI, 
they assume $\sigma_{\mu \rel}=0$ in their plot in Fig.~11. 
We have explicitly confirmed this point by communications 
with the experimental group. We remark that it might not affect 
the results too much due to small number of events at the dip,  
$\delta (N) = \sigma N$.
}
Our results reproduce well the error of $\sin^2{2\theta_{23}}$ 
presented in LOI\@. 
The agreement is particularly good for $\sin^2{2\theta_{23}^\best}=1.0$ 
and is reasonably good for $\sin^2{2\theta_{23}^\best}=0.9$ in the 
interesting region of $\Delta m^2_{31}$.

\section{Combined analysis with the 
LBL appearance or the reactor experiments}

We define $\Delta \chi^2$ for the combined analysis with the 
LBL $\nu_e$ appearance experiment as
\begin{eqnarray}
 \Delta \chi^2_{\mu e} (s_{23}, \Delta m^2_{31}, s_{13})
 &\equiv&
 \Delta \chi^2_\mu (s_{23}, \Delta m^2_{31}, s_{13})
     + \min_\delta
     \Delta \chi^2_e (s_{13} , \delta, s_{23}),\\[3mm]
 \Delta \chi^2_e (s_{13}, \delta, s_{23})
 &\equiv&
  \frac{\left( N_e - N_e^\best \right)^2}
       { N_e^\best + B_e
         + \sigma_{e \rel}^2 (N^\best_e)^2
         + \sigma_{e \BG}^2 B_{e}^2 }
\label{chi2e}.
\end{eqnarray}
The numbers of signal and background events are denoted 
in (\ref{chi2e}) as 
$N_e$ and $B_e$, respectively, which are defined without 
binning for $\nu_e$ appearance. 
In the analyses in this section, the systematic errors 
in disappearance and appearance measurement are assumed 
to be the same, 
$\sigma_{e \rel} = \sigma_{\mu \rel} = 5\%$
and $\sigma_{e \BG} = \sigma_{\mu \BG} = 10\%$. 
In the definition of $\Delta \chi^2_e$,
we fix for simplicity $\theta_{23}$ and $\Delta m^2_{31}$
at their input values 
$\theta_{23}^\best$ and $(\Delta m^2_{31})^\best$,
respectively.
For more details on our treatment of the appearance experiment, 
see \cite{reactorCP}.
 The allowed region is obtained
by projecting three-dimensional manifold of
$\Delta \chi^2_{\mu e} = \mbox{constant}$ onto
the $s_{23}^2 \mbox{-} \Delta m^2_{23}$ plane, 
the similar procedure as we defined in Appendix B for the 
analysis in Sec.~\ref{analysis}.


Here are some technical comments on how we simulate the 
number of events after experimental cut. 
Reconstruction of neutrino energy can be done for quasi-elastic events 
but not for inelastic events unless reaction products deposit all 
the energies and they are detected in the water Cherenkov detector. 
In our analysis, the relation between reconstructed neutrino 
energy $E_\rec$ of background events and true energy $E_\nu$ 
is assumed to be
\begin{eqnarray}
 E_\rec = E_\nu - 0.3\GeV
\end{eqnarray}
by using the comparison between Monte Carlo simulation done by 
the JPARC-SK group. See Fig.~2 in~\cite{JPARC}.

For signal events,
$E_\rec = E_\nu$ is assumed. Then,
from a spectrum $N_{ac}$ calculated by JPARC-SK group~\cite{private}
with a certain set $x_0$ of parameter values,
the spectrum after cut for any values of parameters $x$
is estimated on a bin-by-bin basis as
\begin{eqnarray}
 N_{ac} (x) = \frac{N_{bc} (x)}{N_{bc} (x_0)}\ N_{ac} (x_0),
\end{eqnarray}
where $N_{bc}$ represents the calculated number of events before cut.
 In our analysis,
this prescription is used for all number of events in JPARC-SK.

We define $\Delta \chi^2$ for the combined analysis with reactor 
$\theta_{13}$ experiment via the following considerations: 
The atmospheric $\Delta m_{31}^2$ will be determined by the 
location of the dip in the energy distribution of muons in 
disappearance measurement corresponding to the first 
oscillation maximum. Then, the reactor experiments can determine 
$\theta_{13}$ independent of the other oscillation parameters 
as well as the matter effect as discussed in \cite{MSYIS}. 
Therefore, it acts like giving additional constraint purely on 
$\theta_{13}$.
The appropriate $\Delta \chi^2$ then takes a very simple one, 
\begin{eqnarray}
\Delta \chi^2_{\mu r} (s_{23}, \Delta m^2_{23}, s_{13}) 
\equiv
\Delta \chi^2_{\mu} 
(s_{23}, \Delta m_{31}^2, s_{13}) +
\Delta \chi^2_{r}(s_{13}) .
\end{eqnarray}
For $\Delta \chi^2_{r}$ of reactor experiment
we use an effective parametrization 
\begin{eqnarray}
\Delta \chi^2_{r} = 
\frac{\left(
\sin^2{2\theta_{13}} - \sin^2{2\theta_{13}^\best}
\right)^2}
{\sigma_{r}^2}
\end{eqnarray}
rather than dealing with a particular experimental setup explicitly, 
where $\sin^2{2\theta_{13}^\best}$ is the true value of 
$\sin^2{2\theta_{13}}$ assumed in the analysis. 
Note that the uncertainty in $\sin^2{2\theta_{13}}$
in the reactor-neutrino disappearance measurement is 
almost independent of true value of $\theta_{13}$ as was 
shown in our previous analyses~\cite{MSYIS,reactorCP}.
This feature is in contrast with the one in the LBL
appearance measurement of $\theta_{13}$ mentioned in the 
previous section.  
We take, based on the references, 
$\sigma_{r} = 0.02/\sqrt{2.7}$ so that 
$\delta \left(\sin^2{2\theta_{13}} \right) = 0.02$ at 90\%~CL 
in the analysis with 1 degree of freedom.
The effective $\Delta \chi^2_{r}$ has a definite merit because it 
does not depend on details of the particular experiments, and 
the value of $\sigma_{r}$ is a representative of various different 
analyses done before.

\begin{acknowledgments}
We thank Tsuyoshi Nakaya, Katsuki Hiraide, Takashi Kobayashi, 
and Kenji Kaneyuki for useful informative correspondences. 
Takaaki Kajita and Alexei Smirnov kindly read through earlier 
versions of the manuscript and made illuminating remarks. 
This work was supported by the Grant-in-Aid for Scientific Research
in Priority Areas No. 12047222, Japan Ministry
of Education, Culture, Sports, Science, and Technology, and by 
the Grant-in-Aid for Scientific Research, No. 16340078, 
Japan Society for the Promotion of Science. 
The work of HS was supported by the Research Fellowship
of JSPS for Young Scientists.

\end{acknowledgments}


\end{document}